\documentclass[10pt,aps,pra,showpacs,twocolumn]{revtex4}
\usepackage{amsmath}
\usepackage{amssymb}

\usepackage{amsmath,amssymb}
\usepackage[usenames]{color}
\usepackage{mathbbol}
\usepackage{amssymb}
\usepackage{grffile}
\usepackage[pdftex]{graphicx}
\usepackage{amsmath, amstext, amssymb, amsfonts, amsxtra}
\usepackage{textcomp}
\usepackage{xspace}

\DeclareSymbolFontAlphabet{\amsmathbb}{AMSb}

\newcommand{\nop}{\hat{n}}

\newcommand{\sgz}{\hat{\sigma}^z}
\newcommand{\sgp}{\hat{\sigma}^{+}}
\newcommand{\sgm}{\hat{\sigma}^{-}}
\newcommand{\aop}{\hat{a}}
\newcommand{\adop}{\hat{a}^{\dagger}}
\newcommand{\alop}{\hat{\alpha}}
\newcommand{\aldop}{\hat{\alpha}^{\dagger}}
\newcommand{\bop}{\hat{b}}
\newcommand{\bdop}{\hat{b}^{\dagger}}
\newcommand{\cop}{\hat{c}}
\newcommand{\cdop}{\hat{c}^{\prime}}

\newcommand{\Hop}{\hat{H}}
\newcommand{\Nop}{\mathcal{N}}
\newcommand{\rhoop}{\hat{\rho}}
\newcommand{\hm}{\bold{h}}
\newcommand{\Mm}{\bold{M}}
\newcommand{\Pm}{\bold{P}}
\newcommand{\Pop}{\mathcal{P}}
\newcommand{\PA}{\mathcal{A}}
\newcommand{\PB}{\mathcal{B}}
\newcommand{\PC}{\mathcal{C}}
\newcommand{\Am}{\bold{A}}
\newcommand{\Om}{\bold{O}}
\newcommand{\IM}{\bold{I}}
\newcommand{\Jm}{\bold{J}}
\newcommand{\Cm}{\bold{C}}
\newcommand{\Dm}{\bold{D}}
\newcommand{\Um}{\bold{U}}
\newcommand{\Vm}{\bold{V}}
\newcommand{\Ym}{\bold{Y}}
\newcommand{\Zm}{\bold{Z}}
\newcommand{\Wm}{\bold{W}}
\newcommand{\Qm}{\bold{Q}}
\newcommand{\Km}{\bold{K}}
\newcommand{\Xm}{\bold{X}}

\newcommand{\Sop}{\hat{S}}
\newcommand{\Top}{\hat{T}}

\newcommand{\Dop}{\mathcal{D}}

\newcommand{\Lmp}{\bold{\Lambda}^{+}}
\newcommand{\Lmm}{\bold{\Lambda}^{-}}

\newcommand{\Omegam}{\bold{\Omega}}
\newcommand{\deltam}{\bold{\delta}}
\newcommand{\id}{\bold{1}} 
\newcommand{\zero}{\bold{0}}
\newcommand{\im}{{\rm i}}

\newcommand{\hc}{{\rm H.c.}} 
\newcommand{\diag}{{\rm diag}}
\newcommand{\tr}{{\rm tr}}

\newcommand{\stackvert}[2]{\genfrac{(}{)}{0pt}{}{#1}{#2}}
\newcommand{\lmp}{{\pmb{\lambda}_P}} 
\newcommand{\lmpc}{{\pmb{\lambda}_P^{\ast}}} 
\newcommand{\pare}[1]{\left(#1 \right)}

\newcommand{\sutd}{Singapore University of Technology and Design, 8 Somapah Road, 487372 Singapore}

\begin{document}

\title{Solutions for dissipative quadratic open systems: part II - fermions} 
\author{Chu Guo}
\affiliation{\sutd} 
\author{Dario Poletti}
\affiliation{\sutd} 

\begin{abstract} 
This is the second part of a work in which we show how to solve a large class of Lindblad master equations for non-interacting particles on $L$ sites. Here we concentrate on fermionic particles. In parallel to part I for bosons, but with important differences, we show how to reduce the problem to diagonalizing an $L \times L$ non-Hermitian matrix which, for boundary dissipative driving of a uniform chain, is a tridiagonal bordered Toeplitz matrix. In this way, both for fermionic and spin systems alike, we can obtain analytical expressions for the normal master modes and their relaxation rates (rapidities) and we show how to construct the non-equilibrium steady state.   
\end{abstract}

\date{\today}
\maketitle


\section{Introduction} 

Boundary dissipatively driven systems are quantum systems which are coupled, at their ends, with the environment. Understanding the particle and energy transport in these systems depending on their parameters and the coupling to the environment is a problem which has been attracting a large deserved interest. 
One approach to study such open systems is by using a master equation as derived in \cite{Lindblad1976} and \cite{GoriniSudarshan1976} which is commonly referred to as Lindblad master equation. 

The knowledge of analytical solutions would help gather a deeper intuition of these systems and also it would allow to test the validity of numerical or approximate solutions near the regime of validity of the exact ones. However few models, for dissipative quantum systems, have been solved analytically. 

For fermionic and spin systems we should mention that the non-equilibrium steady state can, in certain cases, be constructed analytically via a matrix product ansatz \cite{Prosen2011a, Prosen2011b, Prosen2014, Popkov2013b, Popkov2014}. For a review of this technique see \cite{Prosen2015}. In \cite{Znidaric2010} the author finds a perturbative expression for the steady state of a boundary driven XX chain with dephasing, and is able to compute exactly one and two-point correlations, using a cleverly designed ansatz. Moreover, in \cite{MedvedyevaProsen2016}, the tight-binding fermionic chain with dephasing is mapped to a Hubbard model with imaginary interaction. The authors are then able to compute, for example, the eigenvalues of the Lindblad master equation using the Bethe Ansatz. 

In \cite{Prosen2008}, the author showed that, analogously to closed systems but in an enlarged space, a quadratic Lindblad master equation for fermions can be written (analogous to Bogoliubov transformation) as a sum of occupation number of so-called normal master modes, times their ``rapidities'', a complex number which characterize their evolution (typically a decay plus phase shift). In this way he was able to convert the problem into solving for the eigenvalues and eigenvectors of a $4L\times 4L$ matrix, where $L$ is the number of sites on the fermionic chain. This allowed him to find, for example, solutions for the homogeneous transverse Ising chain.      

Building on this, in part I, \cite{GuoPoletti2016b}, we obtained analytical expression for the rapidities and normal master modes for a quadratic open boundary driven bosonic systems with a number-conserving, uniform, bulk Hamiltonian. This was possible by reducing the problem of finding the rapidities and normal master modes to solving for the eigenvalues and eigenvectors of a tridiagonal bordered Toeplitz $L\times L$ matrix. For a particular type of parameters combinations, and using the results in \cite{Yueh2005} we could also write explicit analytical solutions. 
In part II we extend these results to fermionic systems (and via Jordan-Wigner transformation \cite{JordanWigner1928, LiebMattis1961} to spin systems). We also show that it is possible to convert the problem into finding eigenvalues and eigenvectors of a tridiagonal $L\times L$ matrix. For the uniform XX model with boundary dissipative driving, this matrix is of tridiagonal bordered Toeplitz form and we are able to write an explicit expression for the normal master modes and rapidities. 

In this article we attempt to be as self-contained as possible while, at the same time, trying to reduce repetitions with what presented in part I. We should stress that imposing fermionic anti-commutation relations between the normal master modes requires a few further steps compared to the bosonic case. Moreover, while the logical flow of the derivations in part I and part II are similar, there are important differences, for example in the sign of various matrices and sub-blocks of matrices. 

This paper is organized as follows: In section \ref{sec:model} we introduce the quadratic fermionic model that we study. In section \ref{sec:diagonalization}, we show how to diagonalize the Lindblad master equation and obtain the normal master modes. In section \ref{sec:exact} we show how to solve analytically the boundary driven XX model. In section \ref{sec:steady_state} we analytically show that a similarity transformation can be used to construct the steady state of the system and in section \ref{sec:conclusion}, we draw our conclusions.

\section{model} \label{sec:model}
We consider an open quantum systems of $L$ sites with fermionic particles. Its dynamics is described by the quantum Lindblad master equation \cite{GoriniSudarshan1976, Lindblad1976}
\begin{align}\label{eq:Lindblad}
\frac{d}{dt} \rhoop = \mathcal{L}(\rhoop) =  -\frac{\im}{\hbar} [ \Hop, \rhoop ] + \mathcal{D}(\rhoop).
\end{align}
$\rhoop$ is the density operator of the system, $\Hop$ is the Hamiltonian, and the dissipator $\mathcal{D}$ models the dissipative part of the evolution. The Hamiltonian $\hat{H}$ is given by 
\begin{align}
\Hop =\sum_{m,n=1}^{L}\hm_{m,n}\aldop_m \alop_n, \label{eq:genHamiltonian}
\end{align}
where $\hm$ is an $L\times L$ Hermitian matrix. The dissipative part is given by
\begin{align}
\mathcal{D}(\rhoop) = \sum_{i,j=1}^{L}&\left[\Lmp_{i,j}(\aldop_{i}\rhoop \alop_{j}-\alop_{j}\aldop_{i}\rhoop) \right. \\ &+ \left.  \Lmm_{i,j}(\alop_{i}\rhoop \aldop_{j}-\aldop_{j}\alop_{i}\rhoop)+ \hc \right],
\end{align}
where $\Lmp$ and $\Lmm$ are $L \times L$ Hermitian and non-negative matrices. The operators $\alop_j$ and $\aldop_j$ respectively annihilate or create a fermion at site $j$.

\section{solving the master equation} \label{sec:diagonalization}
\subsection{Mapping the density operator into new representations}

In order to be able to treat the open fermionic system, we proceed similarly to the bosonic case \cite{GuoPoletti2016b}, however, in order to preserve anti-commutation relations between the operators, we will have to use an ulterior transformation.

First we perform a one-to-one mapping from the density operator basis elements $\vert n_1, n_2, \dots n_L \rangle \langle  n'_1, n'_2, \dots n'_L \vert $ to a state vector basis (with $2L$ sites) which we denote as $\vert n_1, \dots n_L, n'_1, \dots n'_{L} \rangle_{\PA}$ (see for example \cite{ProsenPizorn2008, ProsenZunkovic2010, Prosen2010, PizornTroyer2013}). As a result, the operator $\alop_i$ acting on site $i$ to the left of the density matrix is mapped to $\aop_i$ acting on the state vector on the $i$-th site too, while the operator $\alop_i$ acting on the right of the density matrix is mapped to $\adop_{L+i}$ acting on the state vector. We refer to this new representation defined by the $2L$ modes $\aop$ as $\PA$. The $2L$ modes $\aop_i$ satisfy the following relations
\begin{subequations}
\begin{align}
&\lbrace \aop_i, \aop_j \rbrace = 0, \;\;\;\;\lbrace \adop_i, \adop_j \rbrace = 0 \\
& \lbrace \aop_{L+i}, \aop_{L+j} \rbrace = 0, \;\;\;\;\lbrace \adop_{L+i}, \adop_{L+j} \rbrace = 0 \\
& \lbrace \aop_i, \adop_j \rbrace = \deltam_{ij}, \;\;\;\;\lbrace \aop_{L+i}, \adop_{L+j} \rbrace = \deltam_{ij} \\
& \left[ \aop_i, \aop_{L+j} \right] = 0, \;\;\;\;\left[ \adop_i, \adop_{L+j} \right] = 0 \\
& \left[ \aop_i, \adop_{L+j} \right] = 0, \;\;\;\;\left[ \adop_i, \aop_{L+j} \right] = 0 
\end{align}
\end{subequations}
The operators acting on the group of sites $1 \rightarrow L$ and the group of sites $L+1 \rightarrow 2L$ satisfy fermionic anti-commutation relations among themselves separately. However, the operators between these two groups commute with each other. 

To enforce the fermionic anti-commutation relations over all the sites, we perform a second mapping from $2L$ modes $\aop_i$ to another set of $2L$ modes $\bop_i$, which we refer to as the $\PB$ representation:
\begin{subequations}
\begin{align}
& \bop_i  = \aop_i, \;\;\;\; \bdop_i = \adop_i \\
&\bop_{L+i} = \Pop\aop_{L+i}, \;\;\;\; \bdop_{L+i} = \adop_{L+i}\Pop,   
\end{align}
\end{subequations}
where $\Pop$ is the parity operator defined as
\begin{align}
\Pop = e^{\im \pi \Nop}, \;\;\;\;\Nop = \sum_{j=1}^{2L}\bdop_j\bop_j
\end{align}
We note that $\Pop$ anti-commutes with all the operators $\bop, \bdop$, which means
\begin{align}
\lbrace \Pop, \bop_i \rbrace = 0, \;\;\;\; \lbrace \Pop, \bdop_i \rbrace = 0,
\end{align}
for $1 \leq i \leq 2L$.

Now it is straightforward to verify that the $2L$ modes $\bop$ satisfy the fermionic anti-commutation relations 
\begin{subequations}
\begin{align}
& \lbrace \bop_i, \bop_{L+j} \rbrace = \lbrace \aop_i, \Pop\aop_{L+j}  \rbrace = 0 \\
& \lbrace \bdop_i, \bdop_{L+j} \rbrace = 
\lbrace \adop_i, \adop_{L+j}\Pop  \rbrace = 0 \\
& \lbrace \bop_i, \bdop_{L+j} \rbrace 
= \lbrace \aop_i, \adop_{L+j} \Pop\rbrace = 0 \\
& \lbrace \bdop_i, \bop_{L+j} \rbrace = \lbrace \adop_i, \Pop \aop_{L+j}  \rbrace = 0 \\
& \lbrace \bop_{L+i}, \bop_{L+j} \rbrace = 
\lbrace \Pop \aop_{L+i}, \Pop \aop_{L+j} \rbrace = 0 \\
& \lbrace \bdop_{L+i}, \bdop_{L+j} \rbrace = 
\lbrace \Pop \adop_{L+i}, \adop_{L+j} \Pop\rbrace = 0 \\
& \lbrace \bop_{i}, \bdop_{j} \rbrace = 
\lbrace \Pop \aop_{i}, \adop_{j}\Pop\rbrace = \deltam_{ij} \\ 
& \lbrace \bop_{L+i}, \bdop_{L+j} \rbrace = 
\lbrace \Pop \aop_{L+i}, \adop_{L+j}\Pop\rbrace = \deltam_{ij}
\end{align}
\end{subequations}
The unitary part of Eq.(\ref{eq:Lindblad}) can be written in the $\PB$ representation as
\begin{align}\label{eq:Hnew}
[\Hop, \rhoop]_{\PB}  
 &= \sum_{i,j=1}^L \left( \hm_{ij}\bdop_i \bop_j - \hm_{ji}\bdop_{L+i}\bop_{L+j} \right)\vert \rho  \rangle_{\PB} ,
\end{align}
and the dissipative part of Eq.(\ref{eq:Lindblad}) can be written in the $\PB$ representation as
\begin{align}\label{eq:Dnewinter}
\Dop^{\PB} \vert \rho \rangle_{\PB} = \sum_{i,j=1}^L &\left( \Lmp_{ij}  \bdop_i \bdop_{L+j} \Pop- \Lmp_{ji} \bop_i \bdop_j  + \Lmm_{ji}  \bop_{L+i} \bop_{j} \Pop\right. \nonumber \\
 - &\left. \Lmm_{ji} \bdop_i \bop_j -  {\Lmp_{ij}}^{\ast}  \bdop_{L+i}\bdop_j \Pop- {\Lmp_{ji}}^{\ast}  \bop_{L+i} \bdop_{L+j} \right. \nonumber \\ - 
&\left.{\Lmm_{ji}}^{\ast} \bop_{i}\bop_{L+j} \Pop - {\Lmm_{ji}}^{\ast} \bdop_{L+i}\bop_{L+j} \right)\vert \rho  \rangle_{\PB},  
\end{align}
where $\Dop^{\PB}$ is the dissipator $\Dop$ in the ${\PB}$ representation while $\vert \rho \rangle_{\PB}$ the density operator $\rhoop$ in $\PB$. 

Now the system is almost in quadratic form of operators $\bop_i$, $\bdop_i$ except for the presence of the parity operator $\Pop$. To remove the $\Pop$ operators, we first note that in general we can write $\vert \rho \rangle_{\PB} $ as
\begin{align}\label{eq:generalrhob}
\vert \rho \rangle_{\PB} = \sum_{n_1, \dots, n_L, n_1^{\prime}, \dots, n_L^{\prime}} \bop_1^{\dagger, n_1}\dots\bop_L^{\dagger, n_L}\bop_{L+1}^{\dagger, n_1^{\prime}}\dots\bop_{2L}^{\dagger, n_L^{\prime}} \vert \zero \rangle_{\PB},
\end{align}
where $n_i, n_i^{\prime} = 0, 1$ for $1 \leq i \leq L$, and $\vert \zero \rangle_{\PB}$ is the vacuum state $\vert \zero \rangle \langle \zero \vert$ in the $\PB$ representation. We can see that $\Pop$ conserves the parity of the number of operators of each term in the Eq.(\ref{eq:generalrhob}), which is $N = \sum_{i=1}^L (n_i + n_i^{\prime})$. Therefore, the even sector, defined as the group of terms for which $N$ is even, and the odd sector, defined as the group of terms for which $N$ is odd, of $\vert \rho \rangle_{\PB}$, when acted on by $\Pop$, will obtain opposite signs. Moreover, we can see that each term in Eqs.(\ref{eq:Hnew}, \ref{eq:Dnewinter}) does not change the parity of $N$, which means the even sector and the odd sector of $\vert \rho \rangle_b$ are decoupled under the evolution of Eq.(\ref{eq:Lindblad}). Thus we can treat them separately, and in the following we only consider the even sector for which we can just set $\Pop = 1$ \cite{oddsector}. Hence we have
\begin{align}\label{eq:Dnew}
\Dop^{\PB} \vert \rho \rangle_{\PB} =  \sum_{i,j=1}^L & \left( \Lmp_{ij} \bdop_i \bdop_{L+j} - \Lmp_{ji} \bop_i \bdop_j  + \Lmm_{ji} \bop_{L+i} \bop_{j} -\right. \nonumber \\
&\left. \Lmm_{ji} \bdop_i \bop_j  - {\Lmp_{ij}}^{\ast}  \bdop_{L+i} \bdop_j - {\Lmp_{ji}}^{\ast}  \bop_{L+i} \bdop_{L+j}\right. \nonumber \\ - 
&\left.{\Lmm_{ji}}^{\ast} \bop_{i}\bop_{L+j}  - 
{\Lmm_{ji}}^{\ast} \bdop_{L+i}\bop_{L+j} \right)\vert \rho  \rangle_{\PB} ,
\end{align}

\subsection{The master equation in the new representation}

Combining Eqs.(\ref{eq:Hnew}, \ref{eq:Dnew}), the Linbladian $\mathcal{L}$ of Eq.(\ref{eq:Lindblad}) can be written in the $\PB$ representation as
\begin{align}\label{eq:master}
\mathcal{L}^{\PB} &= \left(
                                                         \begin{array}{cc}
                                                          \textbf{b}_{1\rightarrow L}^{\dagger} \\ \textbf{b}_{L+1\rightarrow 2L} \\
                                                         \end{array}
                                                       \right)^t \Mm
             \left(
                                                         \begin{array}{cc}
                                                          \textbf{b}_{1\rightarrow L} \\
                                                          \textbf{b}_{L+1\rightarrow 2L}^{\dagger} \\
                                                         \end{array}
                                                       \right) \nonumber \\
&-\left(
                                                         \begin{array}{cc}
                                                          \textbf{b}_{1\rightarrow L} \\ \textbf{b}_{L+1\rightarrow 2L}^{\dagger} \\
                                                         \end{array}
                                                       \right)^t \Mm^t
             \left(
                                                         \begin{array}{cc}
                                                          \textbf{b}_{1\rightarrow L}^{\dagger} \\
                                                          \textbf{b}_{L+1\rightarrow 2L} \\
                                                         \end{array}
                                                       \right) \nonumber \\
                                             &-\tr({\Lmm}^t +\Lmp), 
\end{align}
where $\Mm$ is a $2L \times 2L$ matrix,
\begin{eqnarray}
\Mm = \left(
                                                         \begin{array}{cc}
                                                          \Km & \Lmp  \\
                                                          {\Lmm}^t & -\Km^{\dagger} \\
                                                         \end{array}
                                                       \right) 
\end{eqnarray}
and $\Km = (-\im \hm/\hbar + \Lmp - {\Lmm}^t)/2$, where with $\Am^t$ we indicate the transpose of the matrix $\Am$. We have also used the notation $\textbf{b}_{1\rightarrow L} $ to mean a column vector with elements $\bop_1, \bop_2 ,\dots,\bop_L$  and $\textbf{b}^{\dagger}_{1\rightarrow L} $ a column vector with elements $\bdop_{1}, \bdop_{2} ,\dots,\bdop_{L}$ (and similarly for both $\textbf{b}_{L+1\rightarrow 2L}$ and $\textbf{b}^{\dagger}_{L+1\rightarrow 2L}$). We should note here the difference of the last term of Eq.(\ref{eq:master}), $-\tr({\Lmm}^t + \Lmp)$, compared to the bosonic case, $\tr({\Lmm}^t - \Lmp)$ \cite{GuoPoletti2016b}.

\subsection{Normal master modes of the master equation}

In general $\Mm$ is not Hermitian and it cannot always be diagonalized, however in the following we start from the assumption that we know a transformation which can diagonalize $\Mm$ and preserves fermionic anti-commutation relations. This assumption is a posteriori verified in all the cases we considered. This transformation is given by the matrices $\Wm_1$ and $\Wm_2$ as follows  
\begin{eqnarray}
 &&\left(
                                  \begin{array}{cc}
                                  \textbf{b}_{1\rightarrow L} \\
                                 \textbf{b}_{L+1\rightarrow 2L}^{\dagger} \\
                                  \end{array}
                                  \right) = \Wm_1\left(
                                  \begin{array}{cc}
                                  \textbf{c}_{1\rightarrow L} \\
                                  \textbf{c}_{L+1\rightarrow 2L}^{\prime} \\
                                   \end{array} \right) \\
                                  &&\left(
                                  \begin{array}{cc}
                                 \textbf{b}_{1\rightarrow L}^{\dagger} \\
                                 \textbf{b}_{L+1\rightarrow 2L} \\
                                  \end{array}
                                  \right) = \Wm_2\left(
                                  \begin{array}{cc}
                                  \textbf{c}_{1\rightarrow L}^{\prime}  \\
                                 \textbf{c}_{L+1\rightarrow 2L}\\
                                 \end{array} \right),                                                        
\end{eqnarray} 
where as for $\textbf{b}_{1\rightarrow L}$ and $\textbf{b}^{\dagger}_{L\rightarrow L}$, $\textbf{c}_{1\rightarrow L}$ means the column vector made of operators $\cop_1, \cop_2 ,\dots,\cop_L$ and $\textbf{c}^{\prime}_{1\rightarrow L}$ means the column vector made of $\cdop_{1}, \cdop_{2} ,\dots,\cdop_{L}$, and similarly for $\textbf{c}_{L+1\rightarrow 2L}$ and $\textbf{c}^{\prime}_{L+1\rightarrow 2L}$. In the following we refer to the new representation defined by $\cop$ as the $\PC$ representation. Using this transformation we get   
 \begin{align}
\mathcal{L}^{\PC} &= \left(
             \begin{array}{cc}
              \textbf{c}_{1\rightarrow L}^{\prime} \\
              \textbf{c}_{L+1\rightarrow 2L} \\
             \end{array}
             \right)^t \Wm_2^{t} \Mm \Wm_1
             \left(
              \begin{array}{cc}
              \textbf{c}_{1\rightarrow L} \\
           \textbf{c}_{L+1\rightarrow 2L}^{\prime} \\
             \end{array}
              \right) \nonumber \\
&- \left(
              \begin{array}{cc}
              \textbf{c}_{1\rightarrow L} \\
           \textbf{c}_{L+1\rightarrow 2L}^{\prime} \\
             \end{array}
              \right)^t \Wm_1^{t} \Mm^t \Wm_2
             \left(
             \begin{array}{cc}
              \textbf{c}_{1\rightarrow L}^{\prime} \\
              \textbf{c}_{L+1\rightarrow 2L} \\
             \end{array}
             \right) \nonumber \\
             &- \tr({\Lmm}^t + \Lmp),
\end{align}
where $\mathcal{L}^{\PC}$ denotes the Lindbladian $\mathcal{L}$ in the $\PC$ representation. The fermionic anti-commutation relation can be written as $$\left\lbrace \left(
               \begin{array}{cc}
              \textbf{b}_{1\rightarrow L} \\
              \textbf{b}_{L+1\rightarrow 2L}^{\dagger} \\
              \end{array}
              \right), \left(
               \begin{array}{cc}
              \textbf{b}_{1\rightarrow L}^{\dagger} \\
             \textbf{b}_{L+1\rightarrow 2L} \\
             \end{array}
            \right)^{t} \right\rbrace = \id_{2L},$$ and requiring for the fermionic anti-commutation relation to apply also to the $\cop$ we get  
            $$\left\lbrace\left(
             \begin{array}{cc}
             \textbf{c}_{1\rightarrow L} \\
            \textbf{c}_{L+1\rightarrow 2L}^{\prime} \\
             \end{array}
             \right), \left(
            \begin{array}{cc}
           \textbf{c}_{1\rightarrow L}^{\prime} \\
             \textbf{c}_{L+1\rightarrow 2L} \\
             \end{array}
             \right)^{t}\right\rbrace = \id_{2L}$$ and hence
\begin{eqnarray}
\Wm_2 = {\Wm_1^t}^{-1}.
\end{eqnarray}
Here we have used $\id_{l}$ for an identity matrix of size $l$. In the following we also use the matrices
\begin{align} 
\Xm_{L}&=\left(
                                                         \begin{array}{cc}
                                                          0 & \id_{L} \\
                                                          \id_{L} & 0 \\
                                                         \end{array}
                                                       \right)\label{eq:eqX} \\
\Ym_{L}&= -\im \left(
                                                         \begin{array}{cc}
                                                          0 & \id_{L} \\
                                                          -\id_{L} & 0 \\
                                                         \end{array}
                                                       \right) \label{eq:eqY} \\
\Zm_{L}&=\left(
                                                         \begin{array}{cc}
                                                         \id_{L}  & 0 \\
                                                          0 & -\id_{L} \\
                                                         \end{array}
                                                       \right)\label{eq:eqZ}
\end{align}
Matrices in Eqs.(\ref{eq:eqX}, \ref{eq:eqY}, \ref{eq:eqZ}), being given by a tensor product between Pauli matrices and identity, satisfy the relations
$\Zm_{L}^2=\id_{2L}, \;\;\Xm_{L}^2=\id_{2L}, \;\;\Ym_{L}^2 = \id_{2L}, \;\;\Zm_{L}\Xm_{L}=-\Xm_{L}\Zm_{L}= \im\Ym_{L}$.

It follows that 
\begin{align}
\mathcal{L}^{\PC} &= \left(
      \begin{array}{cc}
     \textbf{c}_{1\rightarrow L}^{\prime} \\
     \textbf{c}_{L+1\rightarrow 2L} \\
      \end{array}
      \right)^t \Wm_1^{-1} \Mm \Wm_1
             \left(
      \begin{array}{cc}
     \textbf{c}_{1\rightarrow L} \\
     \textbf{c}_{L+1\rightarrow 2L}^{\prime} \\
       \end{array}
      \right) \nonumber
 \\ &-   \left(
      \begin{array}{cc}
     \textbf{c}_{1\rightarrow L} \\
     \textbf{c}_{L+1\rightarrow 2L}^{\prime} \\
       \end{array}
      \right)^t \Wm_1^{t} \Mm^t {\Wm_1^{t}}^{-1} 
             \left(
      \begin{array}{cc}
     \textbf{c}_{1\rightarrow L}^{\prime} \\
     \textbf{c}_{L+1\rightarrow 2L} \\
      \end{array}
      \right) \nonumber \\
      &-\tr({\Lmm}^t+ \Lmm)
\end{align} 
This implies that the problem of finding the normal modes of the system reduces to finding a $\Wm_1$ such that $\Mm$ can be diagonalized, that is, 
\begin{eqnarray}
\Wm_1^{-1}\Mm \Wm_1 = \diag(\beta_1, \beta_2, \dots, \beta_{2L}), \label{eq:eigeneq}
\end{eqnarray}
where $\diag(\vec{v})$ is a diagonal matrix with the elements of the vector $\vec{v}$ on its diagonal. It is then possible to write the following compact form for $\mathcal{L}^{\PC}$:    
 \begin{align}
\mathcal{L}^{\PC} &= 2\sum_{i=1}^{L}(\beta_{i}\cdop_{i}\cop_{i} - \beta_{L+i}\cdop_{L+i}\cop_{L+i}) \nonumber \\
 &-\sum_{i=1}^{L}(\beta_{i}-\beta_{L+i})-\tr({\Lmm}^t + \Lmp). \label{eq:me_beta}
\end{align}

\subsection{Diagonalizing $\Mm $} 

As we have seen until now, due to the different statistics of fermions and bosons, in order to preserve their commutation or anticommutation, for the bosonic case we diagonalize the matrix $\Zm_L\Mm$ \cite{GuoPoletti2016b}, while for the fermionic case the relevant matrix to be diagonalized is $\Mm$. 
Now we explicitly construct the eigenvalues and eigenvectors of the matrix $\Mm$. Noticing the relation 
\begin{eqnarray}
\Ym_L \Mm \Ym_L = -\Mm^{\dagger} 
\end{eqnarray} 
we find that if $x = \left(
                                                         \begin{array}{cc}
                                                          \textbf{u} \\
                                                          \textbf{v} \\
                                                         \end{array}
                                                       \right)$ is a right eigenvector of $\Mm$ with eigenvalue $\omega$, then $x^{\dagger}\Ym$ is a left eigenvector of $\Mm$ with eigenvalue $-\omega^{\ast}$.
In fact   
\begin{eqnarray}\label{eq:lefteigen}
\Mm x = \omega x \rightarrow x^{\dagger}\Mm^{\dagger} = \omega^{\ast}x^{\dagger} \nonumber \\ \rightarrow
x^{\dagger} \Ym_{L} \Ym_{L} \Mm^{\dagger} \Ym_{L} = \omega^{\ast}x^{\dagger} \Ym_{L} \nonumber \\ 
                               \rightarrow  x^{\dagger} \Ym_{L} \Mm  = -\omega^{\ast}x^{\dagger} \Ym_{L} 
\end{eqnarray}
                                                         
Moreover if $x_1$ is a right eigenvector of $\Mm$ with eigenvalue $\omega_1$, and $x_2$ is a right eigenvector of $\Mm$ with eigenvalue $\omega_2$, then if $\omega_1+\omega_2^{\ast} \neq 0$ then $x_1^{\dagger}\Ym_L x_2 = 0$.
In fact 
\begin{eqnarray}
\Mm x_1 &=& \omega_1 x_1; \nonumber \\
\Mm x_2 &=& \omega_2 x_2, \nonumber 
\end{eqnarray}
then 
\begin{eqnarray}
x_1^{\dagger}\Ym_{L} \Mm  &=& -\omega_1^{\ast} x_1^{\dagger}\Ym_{L}; \nonumber \\
\Mm x_2 &=& \omega_2 x_2; \nonumber \\
\rightarrow (\omega_1^{\ast}+\omega_2)x_1^{\dagger}\Ym_{L} x_2 &=& 0 \nonumber
\end{eqnarray}

Since the eigenvalues of $\Mm$ always appear in pairs, we could list the eigenvalues and the corresponding eigenvectors of $\Mm$ as $\omega_1, \omega_2, \dots, \omega_L, -\omega_{1}^{\ast},\dots, \omega_L^{\ast}$, with the matrix $\Wm_1$ composed in each column by the right eigenvectors $\Wm_1=(\vec{x}_1, \vec{x}_2,\dots, \vec{x}_{2L})$. Then following Eq.(\ref{eq:lefteigen}) we know that $\vec{x}^{\dagger}_{L+j} \Ym_L$ is the left eigenvector of $\Mm$ correponding to $\omega_j$, and $\vec{x}^{\dagger}_j \Ym_L$ is the left eigenvector corresponding to $-\omega_j^{\ast}$, for $1 \leq j \leq L$. Therefore the left eigenvectors of $\Mm$ constitute the matrix $\Xm_L \Wm_1^{\dagger} \Ym_L$. We can now choose to renormalize the right eigenvectors as 
\begin{eqnarray}
\im\Xm_L \Wm_1^{\dagger} \Ym_L \Wm_1 = \Zm_L \Leftrightarrow  \Ym_L \Wm_1^{\dagger} \Ym_L \Wm_1 = -\id_{2L}
\end{eqnarray}
so that we have
\begin{eqnarray}
\Wm_1^{-1} &=& -\Ym_L \Wm_1^{\dagger} \Ym_L, \\
\Wm_2 &=& -\Ym_L \Wm_1^{\ast} \Ym_L.
\end{eqnarray}

At this point we define a new $L\times L$ matrix $\Pm$, which satisfies  
\begin{eqnarray} \label{eq:defineP}
\Pm = \Km - \Lmp = (-\im \hm/\hbar - \Lmp - {\Lmm}^t)/2,
\end{eqnarray}
for which we assume to have the eigendecomposition
\begin{align} \label{eq:eigenP}
\Pm \Wm_P = \Wm_P \lmp,
\end{align}
 where $\Wm_P$ and $\lmp$ are eigenvectors and eigenvalues. Then we find that the $2L\times L$ matrix formed by $\stackvert{\Wm_P}{-\Wm_P}$ constitutes $L$ right eigenvectors of $\Mm$, corresponding to $\lmp$, and the $L\times 2L$ matrix $(\Wm_P^{\dagger}\;\; \Wm_P^{\dagger})$ constitutes $L$ left eigenvectors of $\Mm$, corresponding to $-\lmpc$. 
This can be shown from      
\begin{align}
\Mm \stackvert{\Wm_P}{-\Wm_P} &= \stackvert{\Pm \Wm_P}{-\Pm \Wm_P} = \stackvert{\Wm_P}{-\Wm_P}\lmp \nonumber 
\end{align} 
and 
\begin{align}  
( \Wm_P^{\dagger}, \Wm_P^{\dagger} ) \Mm &= (-\Wm_P^{\dagger} \Pm^{\dagger}, -\Wm_P^{\dagger} \Pm^{\dagger}) \nonumber \\ 
&= -\lmpc(\Wm_P^{\dagger}, \Wm_P^{\dagger}) \nonumber
\end{align} 
By denoting the remaining $L$ right eigenvectors of $\Mm$ as $\stackvert{\Cm}{\Dm}$, where $\Cm,$ $\Dm$ are $L\times L$ matrices, we know that they form the right eigenvectors with eigenvalues $-\lmpc$, which are paired with the left eigenvectors $(\Wm_P^{\dagger}\;,\; \Wm_P^{\dagger})$. Also $(-\Dm^{\dagger}\;,\; \Cm^{\dagger})$ will be the left eigenvectors corresponding the eigenvalues $\lmp$, which are paired with the right eigenvectors $\stackvert{\Wm_P}{-\Wm_P}$. 

Therefore $\Wm_1$ and $\Wm_2$ can be written more explicitly as
\begin{align}
\Wm_1 &= \left(
             \begin{array}{cccc}
              \Wm_P & \Cm  \\
              -\Wm_P & \Dm \\
              \end{array}
         \right), \;\;          \Wm_1^{-1} = \left(
             \begin{array}{cccc}
              -\Dm^{\dagger} & \Cm^{\dagger}  \\
              -\Wm_P^{\dagger} & -\Wm_P^{\dagger} \\
              \end{array}
         \right) ,   \label{eq:W1} \\
\Wm_2 &= \left(
              \begin{array}{cccc}
              -\Dm^{\ast} & -\Wm_P^{\ast}  \\
              \Cm^{\ast} & -\Wm_P^{\ast} \\
              \end{array}
          \right) ,  \;\; \Wm_2^{-1} = \left(
              \begin{array}{cccc}
              \Wm_P^t & -\Wm_P^t  \\
              \Cm^t & \Dm^t \\
              \end{array}
          \right) \label{eq:W2}                                                         
\end{align} 
which means
\begin{subequations}
\begin{align}
\textbf{b}_{1 \rightarrow L} &= \Wm_P \textbf{c}_{1 \rightarrow L} + \Cm \textbf{c}_{L+1 \rightarrow 2L}^{\prime}; \\
\textbf{b}_{L+1 \rightarrow 2L}^{\dagger} &= -\Wm_P \textbf{c}_{1 \rightarrow L} + \Dm \textbf{c}_{L+1 \rightarrow 2L}^{\prime}; \\
\textbf{b}_{1 \rightarrow L}^{\dagger} &= -\Dm^{\ast} \textbf{c}_{1 \rightarrow L}^{\prime} - \Wm_P^{\ast} \textbf{c}_{L+1 \rightarrow 2L}; \\
\textbf{b}_{L+1 \rightarrow 2L} &= \Cm^{\ast} \textbf{c}_{1 \rightarrow L}^{\prime} - \Wm_P^{\ast} \textbf{c}_{L+1 \rightarrow 2L}
\end{align} \label{eq:atob}
\end{subequations}
and the inverse equation
\begin{subequations} 
\begin{align}
\textbf{c}_{1 \rightarrow L} &= -\Dm^{\dagger} \textbf{b}_{1 \rightarrow L} + \Cm^{\dagger} \textbf{b}_{L+1 \rightarrow 2L}^{\dagger}; \\
\textbf{c}_{L+1 \rightarrow 2L}^{\prime} &= -\Wm_P^{\dagger} \textbf{b}_{1 \rightarrow L} - \Wm_P^{\dagger} \textbf{b}_{L+1 \rightarrow 2L}^{\dagger}; \\
 \textbf{c}_{1 \rightarrow L}^{\prime} &= \Wm_P^{t} \textbf{b}_{1 \rightarrow L}^{\dagger} - \Wm_P^{t} \textbf{b}_{L+1 \rightarrow 2L}; \\
  \textbf{c}_{L+1 \rightarrow 2L} &= \Cm^{t} \textbf{b}_{1 \rightarrow L}^{\dagger} + \Dm^{t} \textbf{b}_{L+1 \rightarrow 2L}
\end{align} \label{eq:btoa}    
\end{subequations}
Noticing that $\sum \lambda_{P,i} = \tr(\Pm) = [-\im\;\tr(\hm/\hbar) + \tr({\Lmm}^t-\Lmp) ]/2$ and since the $(\lambda_{P,1}, \dots \lambda_{P,L},\; -\lambda_{P,1}^{\ast}, \dots -\lambda_{P,L}^{\ast})$ correspond to the eigenvalues of $\Mm$, $(\beta_{1},\dots \beta_{ 2L})$, we get the following identity           
\begin{align}
\sum_{i=1}^L(\beta_i - \beta_{L+i}) = \sum_{i=1}^L(\lambda_{P,i} + \lambda_{P,i}^{\ast}) = -\tr(\Lmp + {\Lmm}^t), \label{eq:summingrule}
\end{align}
which exactly cancels the last term in the expression of $\mathcal{L}^{\PC}$ in Eq.(\ref{eq:me_beta}). 
 
We can then write $\mathcal{L}^{\PC}$ as
\begin{align}
\mathcal{L}^{\PC} = 2\sum_{i=1}^{L}\lambda_{P,i}\cdop_{i}\cop_{i} + 2\sum_{i=1}^L \lambda_{P,i}^{\ast}\cdop_{L+i}\cop_{L+i}. \label{eq:Llambdap}
\end{align}     
The state $|\rho_{ss}\rangle_{\PC}$ which annihilates all the operator $\textbf{c}_{1 \rightarrow 2L}$ is the steady state because $\mathcal{L}^c |\rho_{ss}\rangle_c = 0$. The $\cop_i$ are the normal master modes of the Lindblad master equation and the $\lambda_{P,i}$ the rapidities.

\subsection{Computing the expectation value $\langle \aldop_i\alop_j \rangle$}
The expectation value $\langle \aldop_i\alop_j \rangle$ is given by $\tr( \aldop_i\alop_j  \rhoop_{ss})$. In the $\PA$ representation this is written as ${}_{\PA}\langle \textbf{1}\vert \adop_i\aop_j \vert \rho_{ss} \rangle_{\PA}$ where ${}_{\PA}\langle \textbf{1} \vert$ is the transpose of the identity operator in the $\PA$ representation, $ \vert \textbf{1} \rangle_{\PA} = \sum_{i_1, i_2, \dots, i_L} \vert i_1, i_2, \dots, i_L, i_1, i_2, \dots, i_L \rangle_{\PA}$. In the following we will compute this quantity by transforming the $\adop_i$ and $\aop_j$ in the $\PC$ representation, and using the fact that the steady state is the vacuum of the $\cop_i$, i.e. $\vert \rho_{ss}\rangle_{\PA}=\vert \zero \rangle_{\PC}$.

Since for $1\leq j \leq L$, $\aop_j = \bop_j, \adop_j = \bdop_j$, we have $\tr(\rhoop \aldop_i \alop_j)= {}_{\PA}\langle \id \vert \bdop_i \bop_j \vert \zero \rangle_{\PC}$. Using Eqs.(\ref{eq:atob}) we get 
\begin{align}
\bdop_i &= -\sum_{k=1}^L{\Dm}_{i,k}^{\ast}\cdop_k - \sum_{k=1}^L {\Wm_P}_{i,k}^{\ast}\cop_{L+k} \\
\bop_j &= \sum_{k=1}^L{\Wm_P}_{j,k}\cop_k + \sum_{k=1}^L \Cm_{j,k}\cdop_{L+k}
\end{align}
Using this we can write 
\begin{align}
\bdop_i \bop_j &= -\sum_{k,m=1}^L \Dm_{i,k}^{\ast}{\Wm_P}_{\;j,m} \cdop_k \cop_m - \sum_{k,m=1}^L \Dm_{i,k}^{\ast}\Cm_{j,m} \cdop_k \cdop_{L+m} \nonumber \\
 &- \sum_{k,m=1}^L {\Wm_P}_{i,k}^{\ast}{\Wm_P}_{\;j,m} \cop_{L+k}\cop_m \nonumber \\
 & - \sum_{k,m=1}^L {\Wm_P}_{i,k}^{\ast}\Cm_{j,m} \cop_{L+k}\cdop_{L+m} \label{eq:aiaj}
\end{align}
We then show that ${}_{\PA}\langle \textbf{1} \vert$ is annihilated by all the operators $\textbf{c}_{1 \rightarrow 2L}^{\prime}$. From Eq.(\ref{eq:btoa}) we note that 
\begin{align}
&{}_{\PA}\langle \id \vert \cdop_i = \sum_{n_1, \dots, n_L} {}_{\PA}\langle \id \vert \sum_{k=1}^L{\Wm_P^t}_{i,k} \left( \bdop_k - \bop_{L+k} \right) \\
&{}_{\PA}\langle \id \vert \cdop_{L+i} = -\sum_{n_1, \dots, n_L} {}_{\PA}\langle \id \vert \sum_{k=1}^L{\Wm_P^{\dagger}}_{i,k} \left( \bop_k + \bdop_{L+k} \right),
\end{align}
for $1 \leq i \leq L$ and $n_i = 0, 1$. It is thus sufficient to prove that 
\begin{align}
&\sum_{n_1, \dots, n_L}{}_{\PA}\langle \id \vert  \left( \bdop_k - \bop_{L+k} \right) = 0 
\end{align} 
and 
\begin{align}
&\sum_{n_1, \dots, n_L}{}_{\PA}\langle \id \vert \left( \bop_k + \bdop_{L+k} \right) = 0,
\end{align}
for $1 \leq k \leq L$. We have, using the operator $\Nop_k = \sum_{j=1}^k \nop_j$ and considering that $\Pop\vert\id\rangle_{\PA}=\vert\id\rangle_{\PA}$ (there is always an even number of particles in the identity of the $\PA$ representation),     
\begin{align}
&\sum_{n_1, \dots, n_L}{}_{\PA}\langle \id \vert  \left( \bdop_k - \bop_{L+k} \right) \nonumber \\
=&\sum_{n_1, \dots, n_L} {}_{\PA}\langle \id \vert\left(\adop_k - \Pop \aop_{L+k} \right) \nonumber \\ 
=& \sum_{n_1, \dots, n_L } {}_{\PA}\langle \id \vert \adop_k-\sum_{n_1, \dots, n_L } {}_{\PA}\langle \id \vert \aop_{L+k} \nonumber \\ 
=& \sum_{n_1, \dots, n_L } (-1)^{\Nop_{k-1}}n_k {}_{\PA}\langle \dots, 1-n_k, \dots, n_k, \dots \vert \nonumber \\ 
-& \sum_{n_1, \dots, n_L} (-1)^{\Nop_{k-1}}\left( 1-n_{k} \right) {}_{\PA}\langle \dots, n_k,\dots, 1-n_k, \dots \vert \nonumber \\ 
=& \sum_{n_1, \dots, n_L} (-1)^{\Nop_{k-1}} (1-n_k) {}_{\PA}\langle \dots, n_k, \dots, 1-n_k, \dots \vert \nonumber \\ 
-& \sum_{n_1, \dots, n_L} (-1)^{\Nop_{k-1}}\left( 1-n_{k} \right) {}_{\PA}\langle \dots, n_k, \dots, 1-n_k, \dots \vert \nonumber \\ 
=& 0,
\end{align}
and
\begin{align}
&\sum_{n_1, \dots, n_L}{}_{\PA}\langle \id \vert \left( \bop_k + \bdop_{L+k} \right) \nonumber \\
=&\sum_{n_1, \dots, n_L} {}_{\PA}\langle \id \vert\left(\aop_k + \adop_{L+k} \Pop \right) \nonumber \\ 
=& \sum_{n_1, \dots, n_L} {}_{\PA}\langle \id \vert \aop_k -\sum_{n_1, \dots, n_L} {}_{\PA}\langle \id \vert \Pop \adop_{L+k} \nonumber \\ 
=& \sum_{n_1, \dots, n_L} (-1)^{\Nop_{k-1}}(1-n_k) \;\;{}_{\PA}\langle \dots, 1-n_k, \dots, n_k, \dots \vert \nonumber \\ 
 -& \sum_{n_1, \dots, n_L} (-1)^{\Nop_{k-1}}n_{k} \;\;{}_{\PA}\langle \dots, n_k, \dots, 1-n_k, \dots \vert \nonumber \\ 
 =& \sum_{n_1,\dots, n_L} (-1)^{\Nop_{k-1}}n_k \;\;{}_{\PA}\langle \dots, n_k, \dots, 1-n_k, \dots \vert \nonumber \\ 
 -& \sum_{n_1, \dots, n_L} (-1)^{\Nop_{k-1}}n_{k} \;\;{}_{\PA}\langle \dots, n_k, \dots, 1-n_k, \dots \vert \nonumber \\ 
 =& 0. 
\end{align}
In both cases we have changed $n_k\rightarrow 1-n_k$ to get the terms to cancel. Using $\Wm_1^{-1}\;\Wm_1 = \id_{2L}$ and Eq.(\ref{eq:W1}) we can write 
\begin{align}
&\Dm = -\Cm - {\Wm_P^{\dagger}}^{-1} \\
&\Cm = \Wm_P \Qm
\end{align}
where $\Qm$ is a $L \times L$ Hermitian matrix. This allows us to write 
\begin{align}
\Wm_1 = \left(
             \begin{array}{cccc}
              \Wm_P & \Wm_P \Qm  \\
              -\Wm_P & -\Wm_P \Qm- {\Wm_P^{\dagger}}^{-1} \\
              \end{array}
         \right) \label{eq:W1sol}
\end{align}
and, from Eq.(\ref{eq:W1sol}), together with the definition $\Omegam = \Wm_P \Qm \Wm_P^{\dagger}$ we have 
\begin{align}
\Pm \Omegam + \Omegam \Pm^{\dagger} = \Lmp.  \label{eq:pxxp}
\end{align}       

Hence, we find that only the last term of Eq.(\ref{eq:aiaj}) does not vanish and gives 
\begin{align}
{}_{\PA}\langle \textbf{1}\vert \adop_i \aop_j \vert \rho_{ss} \rangle_{\PA} &= -{}_{\PA}\langle \textbf{1}\vert \sum_{k,m=1}^L {\Wm_P}_{i,k}^{\ast}\Cm_{j,m} \cop_{L+k}\cdop_{L+m} \vert \zero \rangle_{\PC} \nonumber \\
 &= - \sum_{k,m=1}^L {\Wm_P}_{i,k}^{\ast}\Cm_{j,m} \deltam_{k,l} 
  = -(\Cm {\Wm_P}^{\dagger})_{j,i} \nonumber \\
  &= -(\Wm_P \Qm \Wm_P^{\dagger})_{j,i} = -\Omegam_{j,i}
\end{align}
The observable matrix $\Om_{i,j} = \tr(\rhoop \aldop_i \alop_j)$ is then given by 
\begin{align}
  \Om  =   -\Omegam^t. \label{eq:OmOmega} 
\end{align}  

\section{Exact solution of a boundary driven XX model}\label{sec:exact}
Here we apply our method to directly obtain the spectrum of the Eq.(\ref{eq:master}) for the boundary driven XX model, which can then be solved analytically in the limit of a long chain. We note that an approximate steady state solution, and exact one- and two-point correlations of XX chain, including also local dephasing, were computed in \cite{Znidaric2010}. The Lindblad equation we consider is  
\begin{align}
\mathcal{L}_{\rm XX}(\rhoop)=-\frac{\im}{\hbar} [ \Hop_{\rm XX}, \rhoop ] + \mathcal{D}_{\rm XX}(\rhoop) \label{eq:masterLC} 
\end{align}
with 
\begin{align}
\Hop_{\rm XX} =  J\sum_{l=1}^{L-1}\left(\sgp_l \sgm_{l+1}+\sgm_{l}\sgp_{l+1} \right) +h_z\sum_{l=1}^L \sgz_l \label{eq:HamLC} 
\end{align}
and 
\begin{align}
\mathcal{D}_{\rm XX}(\rhoop) = \sum_{l=1,L}&\left[\Lambda^+_{l}(2\sgp_{l}\rhoop \sgm_{l}-\{\sgm_{l}\sgp_{l}, \rhoop\}) \right. \\ &+ \left.  \Lambda^-_{l}(2\sgm_{l}\rhoop \sgp_{l}- \{\sgp_{l}\sgm_{l}, \rhoop \}) \right], \label{eq:DissLC}    
\end{align}
First we apply the Jordan-Wigner transformation \cite{JordanWigner1928, LiebMattis1961} to make it a fermionic chain
\begin{align}
\sgp_j &= e^{-i\pi \sum_{k=1}^{j-1}\aldop_k \alop_k}\aldop_j \\
\sgm_j &= e^{i\pi \sum_{k=1}^{j-1}\aldop_k \alop_k}\alop_j \\
\sgz_j &= 2\aldop_j \alop_j - 1
\end{align}
with Hamiltonian   
\begin{align}
\Hop_{F} &= J\sum_{m=1}^{L-1} \left(\aldop_m \alop_{m+1} + \aldop_{m+1}\alop_m \right) \nonumber \\
&+ h_z\sum_{m=1}^L \left( 2 \aldop_m \alop_m-1 \right)
\end{align}
In this case, the non-zero elements of the matrix $\hm$ from Eq.(\ref{eq:genHamiltonian}) are 
\begin{eqnarray}
\hm_{j,j} &=& 2h_z \\
\hm_{j,j+1} &=& \hm_{j+1,j} = J.
\end{eqnarray}
The dissipation instead becomes
\begin{align}
&\Dop(\rhoop) \nonumber \\ =& \sum_{m=1, L} \Lambda_m^+ 
\left(2e^{-i\pi\sum_{k=1}^{m-1}\aldop_k \alop_k}\aldop_m\rhoop 
e^{i\pi\sum_{k=1}^{m-1}\aldop_k\alop_k}\alop_m \right. \nonumber \\
&\left. - \lbrace \alop_m \aldop_m, \rhoop \rbrace \right) \nonumber \\
+ & \sum_{m=1, L} \Lambda_m^- 
\left(2e^{i\pi\sum_{k=1}^{m-1}\aldop_k\alop_k}\alop_m\rhoop 
e^{-i\pi\sum_{k=1}^{m-1}\aldop_k\alop_k}\aldop_m \right. \nonumber \\
 &\left. -\lbrace \aldop_m \alop_m, \rhoop \rbrace \right) \nonumber \\
= & \Lambda_1^+ 
\left(2\aldop_1 \rhoop \alop_1 -\lbrace \alop_1 \aldop_1, \rhoop \rbrace \right) + \Lambda_1^- 
\left(2\alop_1\rhoop 
\aldop_1 -\lbrace \aldop_1 \alop_1, \rhoop \rbrace \right) \nonumber \\
+ & \Lambda_L^+ 
\left(2e^{-i\pi\sum_{k=1}^{L-1}\aldop_k \alop_k}\aldop_L\rhoop 
e^{i\pi\sum_{k=1}^{L-1}\aldop_k \alop_k}\alop_L -\lbrace \alop_L \aldop_L, \rhoop \rbrace \right) \nonumber \\
+ & \Lambda_L^- 
\left(2e^{i\pi\sum_{k=1}^{L-1}\aldop_k \alop_k}\alop_L\rhoop 
e^{-i\pi\sum_{k=1}^{L-1}\aldop_k \alop_k}\aldop_L -\lbrace \aldop_L \alop_L, \rhoop \rbrace \right) \nonumber \\
= & \Lambda_1^+ 
\left(2\aldop_1\rhoop \alop_1 -\lbrace \alop_1 \aldop_1, \rhoop \rbrace \right) + \Lambda_1^- 
\left(2\alop_1\rhoop 
\aldop_1 -\lbrace \aldop_1 \alop_1, \rhoop \rbrace \right)\nonumber \\
+ & \Lambda_L^+ 
\left(2\aldop_L e^{-i\pi\sum_{k=1}^{L}\aldop_k\alop_k}\rhoop 
e^{i\pi\sum_{k=1}^{L}\aldop_k\alop_k}\alop_L -\lbrace \alop_L \aldop_L, \rhoop \rbrace \right) \nonumber \\
+ & \Lambda_L^- 
\left(2\alop_L e^{i\pi\sum_{k=1}^{L}\aldop_k\alop_k}\rhoop 
e^{-i\pi\sum_{k=1}^{L}\aldop_k \alop_k}\aldop_L -\lbrace \aldop_L \alop_L, \rhoop \rbrace \right)
\end{align}
where in the last lines we have included an extra term in the string operator and shifted its position. 
Therefore, in the $\PA$ representation, we have
\begin{align}
\Dop^{\PA} =& \Lambda_1^+ (2\adop_1\adop_{L+1} - \aop_1\adop_1 - \aop_{L+1}\adop_{L+1}) \nonumber \\
+&\Lambda_1^-(2\aop_1\aop_{L+1} - \adop_1\aop_1 - \adop_{L+1}\aop_{L+1}) \nonumber \\
+&\Lambda_L^+(2\adop_L\adop_{2L}\Pop - \aop_{L}\adop_L - \aop_{2L}\adop_{2L}) \nonumber \\
+&\Lambda_L^-(2\aop_L\aop_{2L}\Pop - \adop_L\aop_L - \adop_{2L}\aop_{2L}),
\end{align}
which is, in the $\PB$ representation
\begin{align}
\Dop^{\PB} =& \Lambda_1^+ (2\bdop_1\bdop_{L+1}\Pop - \bop_1\bdop_1 - \bop_{L+1}\bdop_{L+1}) \nonumber \\
+&\Lambda_1^-(-2\bop_1\bop_{L+1}\Pop - \bdop_1\bop_1 - \bdop_{L+1}\bop_{L+1}) \nonumber \\
+&\Lambda_L^+(2\bdop_L\bdop_{2L} - \bop_{L}\bdop_L - \bop_{2L}\bdop_{2L}) \nonumber \\
+&\Lambda_L^-(-2\bop_L\bop_{2L} - \bdop_L\bop_L - \bdop_{2L}\bop_{2L}),
\end{align}
Then, restricting ourselves to the even sector as in Sec.\ref{sec:diagonalization}, we can simply write $\Dop^{\PB}$ as
\begin{align}
\Dop^{\PB} =& \Lambda_1^+ (2\bdop_1\bdop_{L+1} - \bop_1\bdop_1 - \bop_{L+1}\bdop_{L+1}) \nonumber \\
+&\Lambda_1^-(-2\bop_1\bop_{L+1} - \bdop_1\bop_1 - \bdop_{L+1}\bop_{L+1}) \nonumber \\
+&\Lambda_L^+(2\bdop_L\bdop_{2L} - \bop_{L}\bdop_L - \bop_{2L}\bdop_{2L}) \nonumber \\
+&\Lambda_L^-(-2\bop_L\bop_{2L} - \bdop_L\bop_L - \bdop_{2L}\bop_{2L}),
\end{align}
which is exactly the same as Eq.(\ref{eq:Dnew}). Hence we can follow the derivation in Sec.\ref{sec:diagonalization}, and reduce the problem of diagonalizing $\mathcal{L}$ to the eigendecomposition of the matrix $\Pm$ defined in Eq.(\ref{eq:defineP}). 

For convenience we rewrite the four coefficients $\Lambda^a_l$ with four new parameters   
\begin{align}
&\Gamma_1 =  \Lambda_1^- + \Lambda^+_1, \;\; \bar{n}_1 = \frac{\Lambda_1^+}{\Gamma_1} \\
&\Gamma_L =  \Lambda_L^- + \Lambda^+_L, \;\; \bar{n}_L = \frac{\Lambda_L^+}{\Gamma_L}
\end{align}
and then all the non-zero elements of matrix $\Pm$ are
\begin{subequations}
\begin{align}
&\Pm_{1,1} = -\im \frac{h_z}{\hbar} -\frac{\Gamma_1}{2}, \;\; \Pm_{L,L} = -\im \frac{h_z}{\hbar} - \frac{\Gamma_L}{2} \\
&\Pm_{m, m} = -\im \frac{h_z}{\hbar},  \;\;\;\; {\rm for } \; 1 < m < L \\
&\Pm_{m, m+1} = \Pm_{m+1,m} =  -\frac{\im J}{2\hbar},   \;\;\;\; {\rm for} \; 1 \leq m < L   
\end{align}\label{eq:PValues}
\end{subequations}
It thus results in that $\Pm$ is a bordered tridiagonal Toeplitz matrix, whose eigenvalues and eigenvectors can be analytically computed \cite{Yueh2005}, and we can follow the derivation in \cite{GuoPoletti2016b} to get the explicit spectrum under the condition $J^2 = \hbar^2\Gamma_1\Gamma_L$. The only difference is that here there is a constant shift $\im h$ for the diagonal terms of matrix $\Pm$ which was not present in \cite{GuoPoletti2016b}. As a result, the eigenvalues of $\Pm$ are
\begin{align}
\hbar\lambda = -J\sin \pare{\alpha} \sinh \pare{\beta} - \im \left[h_z + J\cos \pare{\alpha} \cosh \pare{\beta}\right]   \label{eq:lambda}
\end{align} 
with
\begin{align}
\alpha &= \frac{k\pi}{L} \label{eq:alpha}, \\
\beta &= \frac{1}{2L}\ln\left(\frac{1+\frac{2 \sqrt{\kappa}}{\kappa + 1}\sin \frac{k \pi}{L}}{1 - \frac{2 \sqrt{\kappa}}{\kappa + 1}\sin \frac{k\pi}{L}} \right), \label{eq:beta}
\end{align}
where $k$ is an integer $1 \leq k < L$ and $\kappa = [J/(\hbar\Gamma_1)]^2$. For more details on the steps from Eq.(\ref{eq:PValues}) to Eqs.(\ref{eq:lambda}-\ref{eq:beta}) see \cite{GuoPoletti2016b}.

\section{Computing the steady state}\label{sec:steady_state}

From Eq.(\ref{eq:Llambdap}) we understand that the steady state of the system is the vacuum of the operators $\cop_j$, that is $|\rho_{ss}\rangle=|\zero \rangle_{\PC}$. This is related to the vacuum of the $\bop_j$, $|\zero \rangle_{\PB} = |\zero \rangle_{\PA}$, by a linear transformation. We can then write 
\begin{align} \label{eq:steadystate}
\vert \rho_{ss} \rangle = \vert \zero \rangle_{\PC} = \Sop^{-1} \vert \zero \rangle_{\PB}.  
\end{align}       

In the following we show how to compute $\Sop$ from $\Wm_1$. First we write $\Sop = e^{\Top}$, where $\Top$ is
 \begin{align}
\Top &= \frac{1}{2} \left(
   \begin{array}{cc}
  \textbf{b}^{\dagger}_{1\rightarrow L} \\
   \textbf{b}_{L+1\rightarrow 2L} \\
   \end{array}
  \right)^t \left(
             \begin{array}{cccc}
              \Um & \Vm  \\
              \IM & \Jm \\
              \end{array}
         \right)
  \left(
      \begin{array}{cc}
   \textbf{b}_{1\rightarrow L} \\
  \textbf{b}^{\dagger}_{L+1\rightarrow 2L} \\
   \end{array}
    \right)
 \nonumber \\ 
 & - \frac{1}{2} \left(
      \begin{array}{cc}
   \textbf{b}_{1\rightarrow L} \\
  \textbf{b}^{\dagger}_{L+1\rightarrow 2L} \\
   \end{array}
    \right)^t \left(
             \begin{array}{cccc}
              \Um^t & \IM^t  \\
              \Vm^t & \Jm^t \\
              \end{array}
         \right)
  \left(
   \begin{array}{cc}
  \textbf{b}^{\dagger}_{1\rightarrow L} \\
   \textbf{b}_{L+1\rightarrow 2L} \\
   \end{array}
  \right) \label{eq:Top}
\end{align}
and where $\Um, \Vm, \IM, \Jm$ are $L\times L$ matrices. It should be noted the presence of a minus sign in the second line of Eq.(\ref{eq:Top}) which is different from the bosonic case. Hereafter we will write
\begin{align}
\Wm = \left(
             \begin{array}{cccc}
              \Um & \Vm  \\
              \IM & \Jm \\
              \end{array}
         \right)
\end{align}
To calculate $e^{\Top}\bdop_j e^{-\Top}$ and $e^{\Top}\bop_{L+j} e^{-\Top}$, we use the relations  
\begin{align}
&\hat{E}  := e^{\hat{T}}\bdop_j e^{-\hat{T}} = \sum_{m=1}^{\infty}\frac{1}{m!}\left[\Top,\bdop_j\right]_m \nonumber \\
&\hat{F} :=  e^{\hat{T}}\bop_{L+j} e^{-\hat{T}} = \sum_{m=1}^{\infty}\frac{1}{m!}\left[\Top,\bop_{L+j}\right]_m  \nonumber
\end{align} 
where the nested commutator is defined recursively as $[\hat{A},\hat{B}]_{m+1} \equiv [\hat{A},[\hat{A},\hat{B}]_{m}] $ with $[\hat{A},\hat{B}]_0 \equiv \hat{B}$. 
This results in 
\begin{align}
\Sop \left(
    \begin{array}{cc}
    \textbf{b}^{\dagger}_{1\rightarrow L} \\ 
    \textbf{b}_{L+1\rightarrow 2L} 
    \end{array}
    \right)^t \Sop^{-1}  
         = \left(
    \begin{array}{cc}
    \textbf{b}_{1\rightarrow L}^{\dagger} \\
     \textbf{b}_{L+1\rightarrow 2L} 
    \end{array}
    \right)^t e^{\Wm }
\end{align}
Similarly we can write 
\begin{eqnarray}
\Sop \left(
      \begin{array}{cc}
   \textbf{b}_{1\rightarrow L} \\
  \textbf{b}^{\dagger}_{L+1\rightarrow 2L} \\
   \end{array}
    \right) \Sop^{-1} = e^{-\Wm} \left(
      \begin{array}{cc}
   \textbf{b}_{1\rightarrow L} \\
  \textbf{b}^{\dagger}_{L+1\rightarrow 2L} \\
   \end{array}
    \right)
\end{eqnarray}
which results in  
\begin{align}
& \Sop \mathcal{L} \Sop^{-1} \nonumber \\ = & \left(
    \begin{array}{cc}
    \textbf{b}_{1\rightarrow L}^{\dagger} \\
     \textbf{b}_{L+1\rightarrow 2L} 
    \end{array}
    \right)^t e^{\Wm} \Mm
           e^{-\Wm}  \left(
     \begin{array}{cc}
     \textbf{b}_{1\rightarrow L} \\
      \textbf{b}^{\dagger}_{L+1\rightarrow 2L} \\
      \end{array}
          \right)  \nonumber \\
- &\left(
     \begin{array}{cc}
     \textbf{b}_{1\rightarrow L} \\
      \textbf{b}^{\dagger}_{L+1\rightarrow 2L} \\
      \end{array}
          \right)^t e^{-\Wm^t} \Mm^t e^{\Wm^t}
             \left(
        \begin{array}{cc}
      \textbf{b}^{\dagger}_{1\rightarrow L} \\
     \textbf{b}_{L+1\rightarrow 2L} \\
    \end{array}
    \right)
   \nonumber \\ 
   - & \tr({\Lmm}^t + \Lmp)    
\end{align}
Thus, as in \cite{GuoPoletti2016b} by setting $\Wm_1 = e^{-\Wm}$, which means
\begin{align} \label{eq:logofeigenspace}
\Wm =  -\log{\Wm_1},
\end{align} 
we can diagonalize $\Sop \mathcal{L} \Sop^{-1}$ whose steady state is $\vert \zero\rangle_{\PB}$. This then allows to derive Eq.(\ref{eq:steadystate}).

\section{Conclusions}\label{sec:conclusion} 

Similarly to part I for bosonic particles, but with important differences, we have shown how to map the problem of computing the relaxation rates and the normal master modes of a Lindblad master equation for non-interacting fermions to the diagonalization of a tridiagonal bordered Toeplitz matrix. We have used this to find analytical solutions for the normal master modes and the rapidities of the Lindblad master equation for a boundary driven spin chain (more precisely of the even sector of the enlarged space $\PB$). This can allow one to effectively study the time evolution of the system and also to have an expression for the steady state too. We should stress further that, while the structure of the analysis parallels that of the bosonic case in part I, imposing anti-commutation relations between the operators in the enlarged space $\PC$ requires further intermediate steps and important differences in the actual matrices used. We would like to highlight that while we have explicitly computed the rapidities and normal master modes for a particular type of XX model, it is possible to recover analytical expressions for a broader set of systems, as long as the complex matrix $\Pm$ can be reduced to any of the tridiagonal bordered matrices which has been solved in, for instance, \cite{Yueh2005, Kouachi2006, Willms2008}. A particularly interesting case would be that of a boundary driven XX model for which the tunnelling parameter on the even bonds is different from that on the odd bonds (where by bond we mean the link between two sites). In this case $\Pm$ would reduce to a matrix of the form given by Eqs.(1,2) in \cite{Kouachi2006}.

\begin{acknowledgments} 
D.P. acknowledges support from Singapore Ministry
of Education, Singapore Academic Research Fund Tier-I (project SUTDT12015005).
\end{acknowledgments}


\begin{thebibliography}{99}
\bibitem{oddsector} The odd sector can be solved following similar steps however it should be noted that i) the unique steady state is in the even sector and ii) the odd sector is irrelevant in the computation of quadratic observables like currents and densities.
\end{thebibliography}
\end{document}